\documentclass[preprint,groupedaddress,amsmath,amssymb,showpacs]{revtex4-1}
\usepackage{graphicx,mathrsfs,epsfig,amsmath,amssymb,multirow,color}
\usepackage{dcolumn}
\usepackage{bm}

\allowdisplaybreaks
\def\be{\begin{equation}}
\def\ee{\end{equation}}
\def\bea{\begin{eqnarray}}
\def\eea{\end{eqnarray}}
\def\bes{\begin{subequations}}
\def\ees{\end{subequations}}

\begin{document}
\title{\LARGE \sf Storage and retrieval of (3+1)-dimensional weak-light bullets and vortices in a coherent atomic gas}
\author{Zhiming Chen$^{1}$, Zhengyang Bai$^{1}$, Hui-jun Li$^{2, 1}$, Chao Hang$^{1}$, and Guoxiang Huang$^{1,\ast}$ }
\affiliation{ $^{1}$State Key Laboratory of Precision Spectroscopy and Department of Physics, East China Normal University, Shanghai 200062, China\\
$^{2}$Institute of Nonlinear Physics and Department of Physics, Zhejiang Normal University, Jinhua 321004, Zhejiang, China\\
$^{\ast}$Correspondence should be addressed to G.H. (gxhuang@phy.ecnu.edu.cn)}


\date{\today}
\maketitle

\noindent
\textbf{A robust light storage and retrieval (LSR) in high dimensions is highly desirable for light and quantum information processing. However, most schemes on LSR realized up to now encounter problems due to not only dissipation, but also dispersion and diffraction, which make LSR with a very low fidelity. Here we propose a scheme to achieve a robust storage and retrieval of weak nonlinear high-dimensional light pulses in a coherent atomic gas via electromagnetically induced transparency. We show that it is available to produce stable (3+1)-dimensional light bullets and vortices, which have very attractive physical property and are suitable to obtain a robust LSR in high dimensions.}


The investigation of light storage and retrieval (LSR), a key technique for realizing optical quantum memory, has received much attention in recent years~\cite{Sim,Lvo,San}. One of important techniques for LSR is electromagnetically induced transparency (EIT)~\cite{Fim}, a quantum interference effect typical occurring in a three-level atomic system interacting with a probe and a control laser fields. The origination of EIT is the existence of dark state, which makes not only the absorption (dissipation) of the probe field largely suppressed but also the LSR possible through an adiabatical manipulation of the control field.

Up to now, nearly all studies on LSR have been carried out in various schemes working in linear regime~\cite{Nov,Bus}. Such schemes are simple but encounter the inevitable problem of pulse spreading due to the existence of dispersion, which may result in a serious distortion for retrieved pulse. Recently, the EIT-based LSR has been generalized to weak nonlinear regime, where the storage and retrieval of a (1+1)-dimensional [(1+1)D] {(i.e., the first `1' refers to one spatial dimension, and the second `1' refers to time)} soliton pulse is suggested~\cite{BHH,CBH}. However, because the (1+1)D soliton pulse is unstable in high dimensions due to the existence of diffraction, such scheme is still not realistic or quite limited. For practical applications of optical quantum memory, a challenged problem is to obtain a light pulse that is robust (i.e., with a high fidelity) during storage and retrieval in (3+1)D.

{Before proceeding, we note that in recent years there is much effort focused on high-dimensional optical solitons due to their rich nonlinear physics and important applications~\cite{Kiv,Malomed2}. Although in recent works~\cite{Li,HH,CH} (3+1)D light bullets and vortices in coherent atomic systems have been studied, the possibility of their storage and retrieval is not explored yet to the best of our knowledge.}

Here we propose an EIT-based new scheme to realize a robust LSR for (3+1)D light pulses in a coherent atomic ensemble working in a free space. Based on Maxwell-Bloch equations governing the evolution of atoms and light field we derive a nonlinear equation controlling the motion of the envelope of a probe field. We show the possibility for obtaining (3+1)D light bullets (or called (3+1)D spatiotemporal optical solitons~\cite{Kiv,Malomed2}) and vortices, which have ultraslow propagating velocity and extremely low generation power. We further show that these high-dimensional light pulses can be stabilized by using the balance between dispersion, diffraction, nonlinearity, and by a far-detuned laser field. We demonstrate that these high-dimensional light pulses can be stored and retrieved very stably by switching off and on a control field.

\begin{figure}
\includegraphics[scale=0.7]{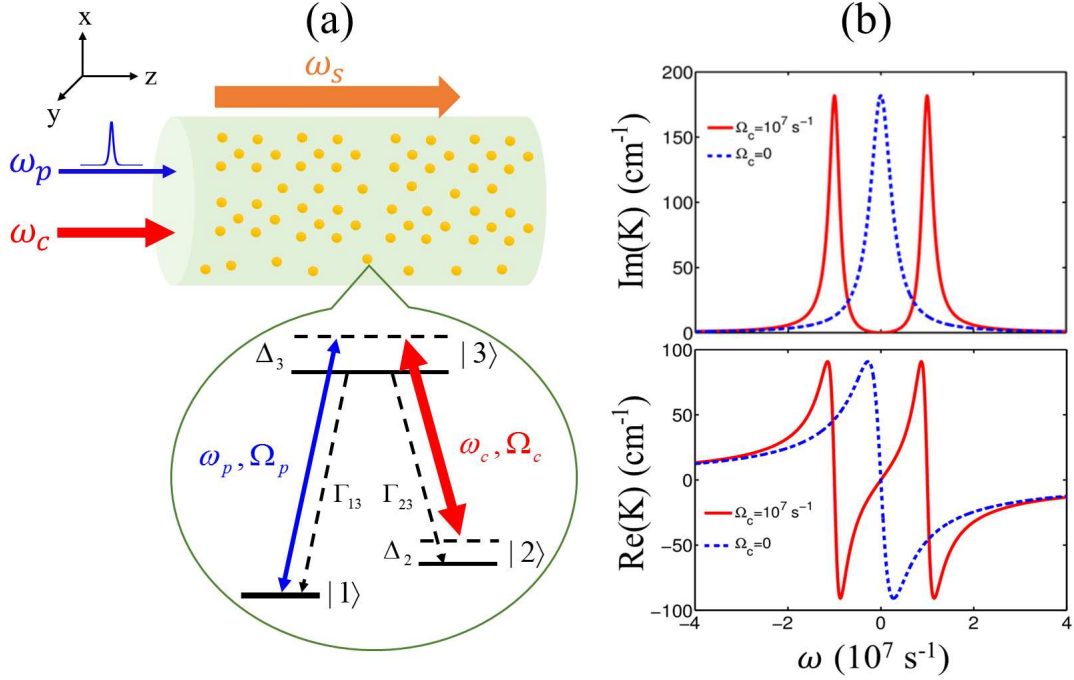}
\caption{\footnotesize {\bf Model and linear dispersion relation.} (a) Possible experimental arrangement of beam geometry. The probe (with angular frequency $\omega_p$ and half Rabi frequency $\Omega_p$) and control (with angular frequency $\omega_c$ and half Rabi frequency $\Omega_c$) fields propagate nearly along $z$ direction. The (orange) thick arrow denotes the Stark field (with angular frequency $\omega_s$) used to stabilize (3+1)D light bullets and vortices. Cold atomic gas are represented by yellow dots. The inset shows the energy-level diagram and excitation scheme of the $\Lambda$-type three-level atoms. $\Delta_2$ and $\Delta_3$ are detunings, $\Gamma_{13}\,(\Gamma_{23})$ is the decay rate from $|3\rangle$ to $|1\rangle$ ($|3\rangle$ to $|2\rangle$). The atoms are initially populated on the ground state $|1\rangle$. (b) The linear dispersion relation $K(\omega)$ of the probe field as a function of $\omega$.}\label{fig:1}
\end{figure}

\vspace{5mm}
\noindent\textbf{\Large \sf Results}\\
\noindent\textbf{Model}.  We consider a cold, lifetime-broadened $\Lambda$-type three-level atomic gas
interacting with a probe field (with pulse length $\tau_0$, center angular frequency $\omega_{p}$, and half Rabi frequency $\Omega_p$) that drives the $|1\rangle \leftrightarrow |3\rangle$ transition, and a continuous-wave control field (with the center angular frequency $\omega_{c}$ and half Rabi frequency $\Omega_c$) that drives
$|2\rangle \leftrightarrow |3\rangle$ transition; see the inset of Fig.~\ref{fig:1}(a).
%
%

For simplicity, we assume the electric field propagates along $z$ direction with the form $\textbf{E}=\sum_{l=p,c}\textbf{e}_l{\cal E}_l e^{i(k_l z-\omega_l t)}+\textrm{c.c.}$, where $\textbf{e}_l$ (${\cal E}_l$) is the unit polarization vector (envelope). A far-detuned laser field (Stark field) used to stabilize (3+1)D light bullets and vortices (see below)
is applied to the system [see Fig.~\ref{fig:1}(a)] with the form
$\textbf{E}_{\textrm{Stark}}(x,y,t) = \textbf{e}_s\sqrt{2} E_s(x,y) \cos(\omega_st)$,
where $\textbf{e}_s$, $E_s$, and $\omega_s$ are the unit polarization vector, field amplitude, and angular frequency, respectively. Due to the existence of the Stark field, an energy shift for the level $|j\rangle$ occurs, i.e., $\Delta E_{j,\textrm{Stark}}=-\alpha_j\langle
\textbf{E}_{\textrm{Stark}}^2\rangle_t/2=-\alpha_j|E_s(x,y)|^2/2$. Here
$\alpha_j$ is the scalar polarizability of the level $|j\rangle$, and $\langle \cdots
\rangle$ denotes the time average in one oscillating cycle.

Under electric-dipole and rotating-wave approximations, the Hamiltonian of the system
in the interaction picture reads
$\hat{\mathcal{H}}_{\textrm{int}}=
-\sum_{j=1}^{3}\hbar\Delta_j^\prime|j\rangle\langle
j|-\hbar\left[\Omega_{p}|3\rangle\langle
1|+\Omega_{c}|3\rangle\langle 2|+\textrm{H.c.}\right]$, with
$\Delta_j^\prime=\Delta_j+[\alpha_j/(2\hbar)]|E_s|^2$, $\Omega_{p}=({\textbf
e}_{p}\cdot{\textbf p}_{13}){\cal E}_{p}/\hbar$, and $\Omega_{c}=({\textbf e}_c\cdot
{\textbf p}_{23}){\cal E}_c/\hbar$. Here
$\Delta_2=\omega_{p}-\omega_{c}-\omega_{21}$ and
$\Delta_3=\omega_{p}-\omega_{31}$ are respectively the two- and one-photon detunings,
${\textbf p}_{jl}$ is the electric-dipole matrix element related to the levels $|j\rangle$ and $|l\rangle$, $\hbar\omega_{jl}=E_j-E_l$ is the energy difference between the level
$|j\rangle$ and the level $|l\rangle$ with $E_j$ the eigenenergy of the level $|j\rangle$.

The equation of motion for density matrix $\sigma$ in the interaction picture reads
\begin{eqnarray}\label{DME}
&&\left(\frac{\partial}{\partial t}+\Gamma\right)\sigma=-\frac{i}{\hbar}\left[\hat{\mathcal{H}}_{\textrm{int}}, \sigma\right],
\end{eqnarray}
where $\sigma$ is a $3 \times 3$ density matrix, $\Gamma$ is a $3 \times 3$ relaxation matrix denoting the spontaneous emission and dephasing. The explicit expressions of Eq.~(\ref{DME}) are presented in Methods.

The equation of motion for $\Omega_p$ can be obtained by the Maxwell equation $\triangledown^2
\textbf{E}-(1/c^2)\partial^2 \textbf{E}/\partial t^2=[1/(\epsilon_0c^2)]\partial^2\textbf{P}/\partial t^2$, {where $\textbf{P}={\cal N}_a\{\textbf{p}_{13}\sigma_{31}\exp[i(k_p z-\omega_p t)]+\textbf{p}_{23}\sigma_{32}\exp[i(k_c z-\omega_c t)]+\textrm{c.c.}\}$ with $\mathcal{N}_a$ the atomic concentration.} Under slowly varying envelope approximation, the Maxwell equation is reduced to~\cite{HDP}
\begin{eqnarray}\label{MEa}
&&i\left(\frac{\partial}{\partial
z}+\frac{1}{c}\frac{\partial}{\partial
t}\right)\Omega_{p}+\frac{c}{2\omega_{p}}\left(\frac{\partial
^2}{\partial x^2}+\frac{\partial
^2}{\partial y^2}\right)\Omega_{p}+\kappa_{13}\sigma_{31}=0,
\end{eqnarray}
where $\kappa_{13}={\cal N}_a\omega_{p}|\textbf p_{13}\cdot \textbf
e_{p}|^2/(2\epsilon_0c\hbar)$, with $c$ the light speed in vacuum.

Our model can be realized by selecting realistic physical systems. One of them is the ultracold $^{87}$Rb atomic gas with the energy levels selected as $|1\rangle=|5^{2}S_{1/2}, F=1\rangle$, $|2\rangle=|5^{2}S_{1/2}, F=2\rangle$, and $|3\rangle=|5^{2}P_{1/2}, F=2\rangle$, respectively. The decay rates are given by
$\Gamma_2\simeq 2\pi \times 1.0$ kHz, and $\Gamma_3\simeq 2\pi \times 5.75$ MHz,
and $\textbf{p}_{13}\simeq\textbf{p}_{23}=2.54\times10^{-27}$ C cm~\cite{Steck}.
If atomic density $\mathcal{N}_a=1.1\times10^{11}$ cm$^{-3}$, $\kappa_{13}$ takes
the value of $3.0\times10^{9}$ cm$^{-1}$ s$^{-1}$.

\vspace{5mm}
\noindent\textbf{Nonlinear envelope equation}{\label{Sec:3a}}.
We use the standard method of multiple scales developed for EIT system~\cite{HDP}
to derive the nonlinear envelope equation for the probe field based on the asymptotic expansion of the
Maxwell-Bloch (MB) Eqs.~(\ref{DME}) and (\ref{MEa})(see Methods).

The first-order solution of the asymptotic expansion reads $\Omega_{p}^{(1)}=Fe^{i\theta}$,
$\sigma_{j1}^{(1)}=\{[\delta_{j3}(\omega+\Delta_2+i\gamma_{21})-\delta_{j2}\Omega_c^\ast]/D\}Fe^{i\theta}$. Here $D=|\Omega_c|^2-(\omega+\Delta_2+i\gamma_{21})(\omega+\Delta_3+i\gamma_{31})$ and
$\theta=K(\omega)z_0-\omega t_0$, with $K(\omega)=\omega/c+\kappa_{13}(\omega+\Delta_2+i\gamma_{21})/D$ (linear dispersion relation). {Note that the frequency and wave number of the probe field are respectively given by $\omega_{p}+\omega$ and $k_{p}+K(\omega)$, so $\omega=0$ corresponds to the center frequency of probe field.}

Fig.~\ref{fig:1}(b) shows the imaginary and real parts of $K(\omega)$, i.e., Im($K$) and Re($K$). The dashed and solid lines are for $\Omega_c=0$ and for $\Omega_c=1.0\times10^7$ s$^{-1}$, respectively. From the upper panel we see that for $\Omega_c=0$ (with no EIT) the probe pulse suffers a large absorption (the dashed line), whereas for $\Omega_c=1.0\times10^7$ s$^{-1}$  (with EIT) a transparency window opens  and hence the probe pulse is nearly free of absorption (the solid line). The lower panel of Fig.~\ref{fig:1}(b) shows the drastic change of dispersion due to EIT, which results in a significant reduction of the group velocity of the probe pulse.

The solvability condition at the second order of the asymptotic expansion is  $i[\partial F/\partial z_{1}+(\partial K/\partial \omega)\partial F/\partial t_{1}]=0$, which means that the probe-pulse envelope $F$ travels with the group velocity $V_g=(\partial K/\partial \omega)^{-1}$. {The nonlinear envelope equation for $F$ is obtained from the solvability condition at the third order, i.e.,
\begin{eqnarray}\label{NLSE2}
i\frac{\partial F}{\partial
z_2}-\frac{1}{2}\frac{\partial^2 K}{\partial \omega^2}\frac{\partial^2 F}{\partial t_1^2}+\frac{c}{2\omega_{p}}\left(\frac{\partial ^2}{\partial
x_1^2}+\frac{\partial ^2}{\partial
y_1^2}\right)F+W_{11}|F|^2Fe^{-2\bar{a}z_2}+W_{12}|E_s^{(1)}|^2F=0,
\end{eqnarray}
where $W_{11}$ is the self-phase modulation coefficient of the probe field and $W_{12}$ is the cross-phase modulation coefficient contributed by the Stark field. The explicit expressions of $W_{11}$ and $W_{12}$ are given in Methods.}

Combining the solvability conditions (i.e., the equations for $F$) at the all orders, we obtain the
unified equation for $F$, which can be written into the dimensionless form
\begin{eqnarray}\label{NLSE3}
i\frac{\partial u}{\partial
s}+\frac{1}{2}\left(g_1\frac{\partial^2 }{\partial \tau^2}+\frac{\partial ^2}{\partial
\xi^2}+\frac{\partial ^2}{\partial
\eta^2}\right)u+g_2|u|^2u+g_3V(\xi,\eta)u=0,
\end{eqnarray}
with $u=\epsilon F/U_0$, $s=z/L_{\rm Diff}$, $\tau=[t-z/{\rm Re}(V_g)]/\tau_0$, $(\xi,\eta)=(x,y)/R$, $g_1=L_{\rm Diff}/L_{\rm Disp}$, $g_2=L_{\rm Diff}/L_{\rm Nonl}$, $g_3=L_{\rm Diff}{\rm Re}(W_{12})E_{0}^2$, and $V(\xi,\eta)=[E_s(\xi,\eta)/E_{0}]^2$. Here $U_0$, $R$, and $E_{0}$ are respectively the typical Rabi frequency,
beam radius, and field amplitude; $L_{\rm Diff}=\omega_pR^2/c$, $L_{\rm Disp}=-\tau_0^2/{\rm Re}(\partial^2 K/\partial \omega^2)$, and $L_{\rm Nonl}=1/[{\rm Re}(W_{11})U_0^2]$ are respectively typical diffraction length, dispersion length, and nonlinear length.

Note that the envelope equation (\ref{NLSE3}) includes dispersion, diffraction, nonlinearity, and ``external'' potential. When obtaining Eq.~(\ref{NLSE3}) we have neglected the imaginary parts of
$\partial^j K/\partial \omega^j$ ($j=1,\,2$), $W_{11}$, and $W_{12}$. This is reasonable because the system works under the EIT condition  $|\Omega_c|^2\gg \gamma_{21}\gamma_{31}$ so that their imaginary parts are much smaller than their real parts. In addition, the diffraction, dispersion, and nonlinearity are assumed to be balanced, i.e., $L_{\rm Diff}=L_{\rm Disp}=L_{\textrm{Nonl}}$, which can be achieved by taking $\tau_0=\sqrt{-{\rm Re}(\partial^2 K/\partial \omega^2)\omega_p/c}R$ and $U_0=\sqrt{c/[\omega_pR^2{\rm Re}(W_{11})]}$ and hence we have $g_1=g_2=1$ in
Eq.~(\ref{NLSE3}). By taking $E_{0}=1/\sqrt{L_{\rm Diff}{\rm Re}(W_{12})}$ we also have $g_3=1$. The ``external''
potential $V(\xi,\eta)$ in Eq.~(\ref{NLSE3}) comes from the Stark field, which can be adjusted and hence useful to control the stability of the light bullets and vortices.

By choosing the realistic system parameters $\Omega_c=9.0\times10^7$ Hz, $\Delta_2=-6.0\times10^6$ Hz, $\Delta_3=-2.0\times10^8$ Hz, $R=40$ $\mu$m, $\tau_0=2.0\times10^{-7}$ s, $U_0=2.87\times10^7$ Hz, and $E_{0}=3.04\times 10^4$ V/cm, we have $L_{\rm Diff}\approx L_{\rm Disp}\approx L_{\textrm{Nonl}}=1.26$ cm, and
\begin{equation}
{\rm Re}(V_g)\approx6.5\times10^{-5}\,c.
\end{equation}
We see that the probe pulse propagates with an ultraslow group velocity.

\vspace{5mm}
\noindent\textbf{Solutions of (3+1)D weak-light bullets and vortices}{\label{Sec:3b}}.
In order to obtain high-dimensional nonlinear localized solutions of the system, we assume
the Stark field has the form of Bessel function, i.e., $E_s(\xi,\eta)=E_{s0}J_l(\sqrt{2b}r)$
($E_{s0}$ and $b$ are real constants; $l$ is an integer; $r=\sqrt{\xi^2+\eta^2}$).
Then Eq.~(\ref{NLSE3}) becomes
\begin{eqnarray}\label{NLSE4}
i\frac{\partial u}{\partial
s}+\frac{1}{2}\left(\frac{\partial^2 }{\partial \tau^2}+\frac{\partial ^2}{\partial
\xi^2}+\frac{\partial ^2}{\partial
\eta^2}\right)u+|u|^2u+v_0^2[J_l(\sqrt{2b}r)]^2u=0,
\end{eqnarray}
with $v_0=E_{s0}/E_{0}$. {Note that Eq.~(\ref{NLSE4}) is similar to that obtained in Ref.~\cite{Malomed3}. However, the physics here is different from that in Ref.~~\cite{Malomed3} because Eq.~(\ref{NLSE4}) describes the nonlinear evolution of the probe-field envelope in the EIT system whereas the equation in Ref.~\cite{Malomed3} governs the dynamics of a Bose-Einstein condensate.} Using the transformation $u=\psi \exp(i\mu s)$, Eq.~(\ref{NLSE4}) is
reduced into $\frac{1}{2}\left(\frac{\partial^2 }{\partial \tau^2}+\frac{\partial ^2}{\partial
\xi^2}+\frac{\partial ^2}{\partial
\eta^2}\right)\psi+|\psi|^2\psi+v_0^2[J_l(\sqrt{2b}r)]^2\psi=\mu \psi$, where
$\mu$ is a propagation constant.

\begin{figure}
\includegraphics[scale=0.78]{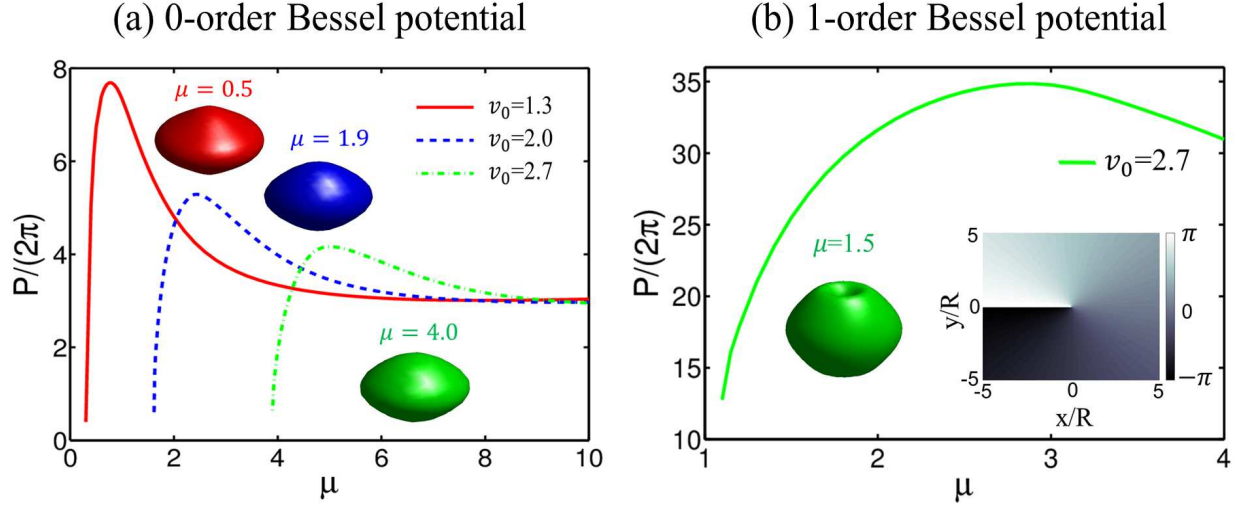}
\caption{\footnotesize {\bf Solutions of (3+1)D light bullets and vortices.} (a) Probe-field power $P$
of several light bullet solutions as functions of $\mu$ and $v_0$, with the Stark field chosen as the zero-order Bessel function (i.e., $l=0$). The solid, dashed, and dotted-dashed lines are for $v_0=1.3,\,2.0,\, \textrm{and} \,2.7$, respectively. Insets give isosurface ($|\psi|=0.05$) plots of light bullets for ($v_0=1.3,\,\mu=0.5$; the red one), ($v_0=2.0,\,\mu=1.9$; the blue one), and ($v_0=2.7,\,\mu=4.0$; the green one), respectively. (b) Probe-field power $P$ of the light vortex solution as a function of $\mu$ and $v_0$, with the Stark field chosen as the first-order Bessel function (i.e., $l=1$). The green solid line is for $v_0=2.7$. Insets display respectively plots of the isosurface ($|\psi|=0.05$) for ($v_0=2.7,\,\mu=1.5$) of the light vortex  and its phase distribution in $x$-$y$ plane.}\label{fig:2}
\end{figure}

Fig.~\ref{fig:2}
%
%
shows the power of the probe pulse defined by $P=2\pi\int\int\int_{-\infty}^{+\infty}|\psi|^2d\xi d\eta d\tau$, which is a function of the propagation constant $\mu$ and the potential strength constant $v_0$. Based on the modified squared-operator method~\cite{Yang}, (3+1)D light bullet solutions are found numerically. Presented in Fig.~\ref{fig:2}(a) is the result of several light bullet solutions for the potential parameters $l=0$ (i.e., the zeroth-order Bessel function) and $b=1$. We see that for different values of $v_0$ ($v_0=1.3$, 2.0, 2.7), $P$ always increases to a maximum firstly, and then
decreases. The stability domain of the light bullet solutions is the one
with $dP/d\mu>0$ according to Vakhitov-Kolokolov (VK) criterion (see~\cite{Yang}), which has been confirmed numerically by using a propagation method. The isosurfaces ($|\psi|=0.05$) of stable light bullet solutions for
($v_0=1.3,\,\mu=0.5$) (the red one), ($v_0=2.0,\,\mu=1.9$) (the
blue one), and ($v_0=2.7,\,\mu=4.0$) (the green one) have been plotted in the figure.

Fig.~\ref{fig:2}(b) shows the result of a light vortex solution for the potential parameters $l=1$
(i.e., the first-order Bessel function) and $b=1$. The light vortex solution with quantum number of
orbital angular momentum  $m=1$ is found. Because the stability domain of the vortex solution
cannot be obtained by the VK criterion, a propagation method is used to study its stability.
The isosurface ($|\psi|=0.05$) and phase distribution of the vortex solution for
($v_0=2.7,\,\mu=1.5$) are plotted in the figure. We found that the vortex solution is fairly
stable during propagation in the region where $dP/d\mu>0$.

The threshold of the optical power density $\bar{P}_{\rm max}$ for producing the (3+1)D light bullets
and vortices given above can be estimated by using Poynting's vector~\cite{HDP}. For light bullets
we obtain
\begin{eqnarray}
&&\bar{P}_{\rm max}\approx 3.77\times10^{-7} \,W.
\end{eqnarray}
A similar conclusion is also obtained for light vortices. Consequently, to produce (3+1)D
light bullets and vortices in the present system very low generation power is needed. {This is drastically contrast to conventional optical media, such as glass-based optical fibers, where generation power at order of kilowatts or even larger is usually needed to
produce light bullets and vortices~\cite{LQW}.}

\vspace{5mm}
\noindent\textbf{Storage and retrieval of (3+1)D light solitons and vortices}.
The principle of EIT-based LSR is well known~\cite{FL}. When switching on the
control field, probe pulse propagates in the atomic medium with
nearly vanishing absorption; by slowly switching off the control
field the probe pulse disappears and gets stored in the form of
atomic coherence; when the control field is switched on again the
probe pulse reappears. However, this principle is usually applied for
linear optical pulses, which may suffer serious distortion due to
the dispersion and/or diffraction. In the following we show
that it is available to realize the LSR of the (3+1)D light bullets and vortices in
our present system.

To this end, we consider the solution of the MB Eqs.~(\ref{DME})
and~(\ref{MEa}) by using a control field that is adiabatically
changed with time $t$ to realize the function of its turning on
and off.  The switching-on and switching-off of the control
field is modeled by the following function
\begin{eqnarray}\label{SWITCH}
\Omega_c=\Omega_{c0}\left\{1-\frac{1}{2}\tanh\left[\frac{t-T_{\textrm{off}}}{T_s}\right]+\frac{1}{2}
\tanh\left[\frac{t-T_{\textrm{on}}}{T_s}\right]\right\},
\end{eqnarray}
where $T_{\textrm{off}}$ and $T_{\textrm{on}}$ are respectively
the times of switching-off and the switching-on of the control
field with a switching time $T_s$. The
storage time of the light bullets and vortices is
approximately given by $T_{\textrm{on}}-T_{\textrm{off}}$.

\begin{figure}
\includegraphics[scale=0.78]{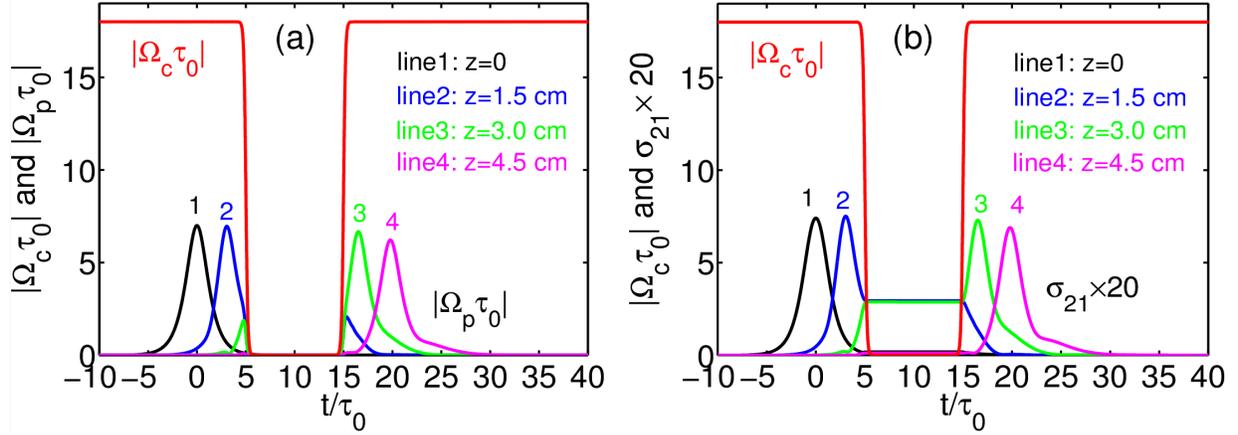}
\caption{\footnotesize {\bf Storage and retrieval of (1+1)D soliton pulse.}
(a) Evolution of $|\Omega_p\tau_0|$ and (b) atomic coherence $\sigma_{21}$ as functions of $z$ and $t$. For a better visualization, $\sigma_{21}$ has been amplified 20 times. Lines 1 to 4 are for $z$=0, 1.5, 3.0, and 4.5 cm, respectively. The control field $|\Omega_c\tau_0|$ is shown in the upper part of each panel.}\label{fig:3}
\end{figure}

We first consider the LSR of the (1+1)D soliton pulse, corresponding the case $\partial^2/\partial\xi^2=\partial^2/\partial\eta^2=0$ and $g_3=0$ in Eq.~(\ref{NLSE3}).
The result of numerical simulation on the time evolution of
$|\Omega_p\tau_0|$ and atomic coherence $\sigma_{21}$ as functions of $z$ and $t$ is presented in Fig.~\ref{fig:3}. The red solid line shown in the upper part of each panel represents the control field $|\Omega_c\tau_0|$. Here we choose $T_s/\tau_0=0.2$, $T_{\textrm{off}}/\tau_0=5.0$, $T_{\textrm{on}}/\tau_0=15.0$, and the other system parameters are mentioned above. The wave shape of the input probe pulse is taken as a hyperbolic secant one, i.e., $\Omega_p(0,t)=7.0\,\textrm{sech}(t/\tau_0)$. Lines 1 to 4 are for propagation distance $z$=0, 1.5, 3.0, and 4.5 cm, respectively.

Shown in Fig.~\ref{fig:3}(a) is the result of $|\Omega_p\tau_0|$. We see that the retrieved pulse has nearly the same shape with the one before the storage. The physical reason of the shape-preservation of the probe pulse before and after the storage is due to a balance between dispersion and nonlinearity, i.e., the pulse is indeed a soliton that is rather stable during the storage and retrieval. Fig.~\ref{fig:3}(b) shows the atomic coherence $\sigma_{21}$, which has been amplified by 20 times for a
better visualization. We see that $\sigma_{21}$ is nonzero during the switch-off of the control field, which is a manifestation of the information transfer (i.e., storage) from the light field to the atomic ensemble.

%

\begin{figure}
\includegraphics[scale=0.75]{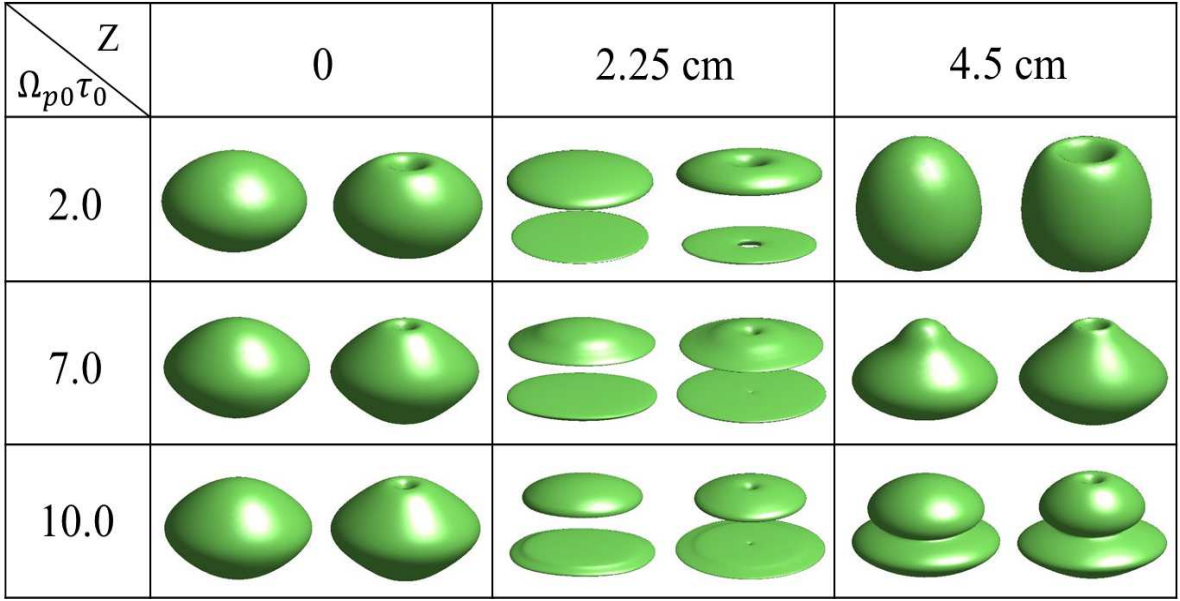}
\caption{\footnotesize {\bf Storage and retrieval of (3+1)D light pulses with different probe-field intensities.} Storage and retrieval of (3+1)D light pulses with the Stark field taken to be the zero-order Bessel function (the left side of each column) and the (3+1)D light pulses with the Stark field taken to be the first-order Bessel function (the right side of each column) for different probe-field intensities (i.e., $\Omega_{p0}\tau_0=2.0,\,7.0,\,10.0$)  at $z=0$ (before the storage), $z=2.25$ cm (during the storage), and $z=4.5$ cm (after the storage), respectively. The second line corresponds to  the storage and retrieval of stable light bullets and vortices.
All figures are isosurface plots with $|\Omega_p\tau_0|=0.5$.}\label{fig:4}
\end{figure}

We now turn to investigate the LSR of the (3+1)D light pulses. Fig.~\ref{fig:4}
shows the storage and retrieval of the light pulses with the Stark field taken to be the zero-order
Bessel function (the left side of each column) and the light pulses with the Stark field taken to be the first-order Bessel function (the right side of each column) for different probe-field intensities,
with the other parameters are the same as used above. Isosurfaces ($|\Omega_p\tau_0|=0.5$) for $\Omega_{p0}\tau_0=2.0$, 7.0, 10.0 at $z=0$ (before the storage), 2.25 cm (during the storage), and 4.5 cm (after the storage)
are illustrated, respectively. The results are the following: (i) For the case of weak probe-field
intensity (the first line in the figure), the probe pulse broadens before and after the storage;
(ii) For the case of moderate probe-field intensity (the second line in the figure), the retrieved
probe pulse has nearly the same shape with the one before the storage; (iii) For the case of strong
probe-field intensity (the third line in the figure), the retrieved probe pulse displays a serious distortion after the storage. From these results, we conclude that in the regime of the moderate
probe-field intensity the storage and retrieval of (3+1)D light pulses are robust, which is desirable
for light and quantum information processing in high dimensions. This regime is just
the one where stable light bullets and vortices can form.
%
%

\begin{figure}
\includegraphics[scale=0.75]{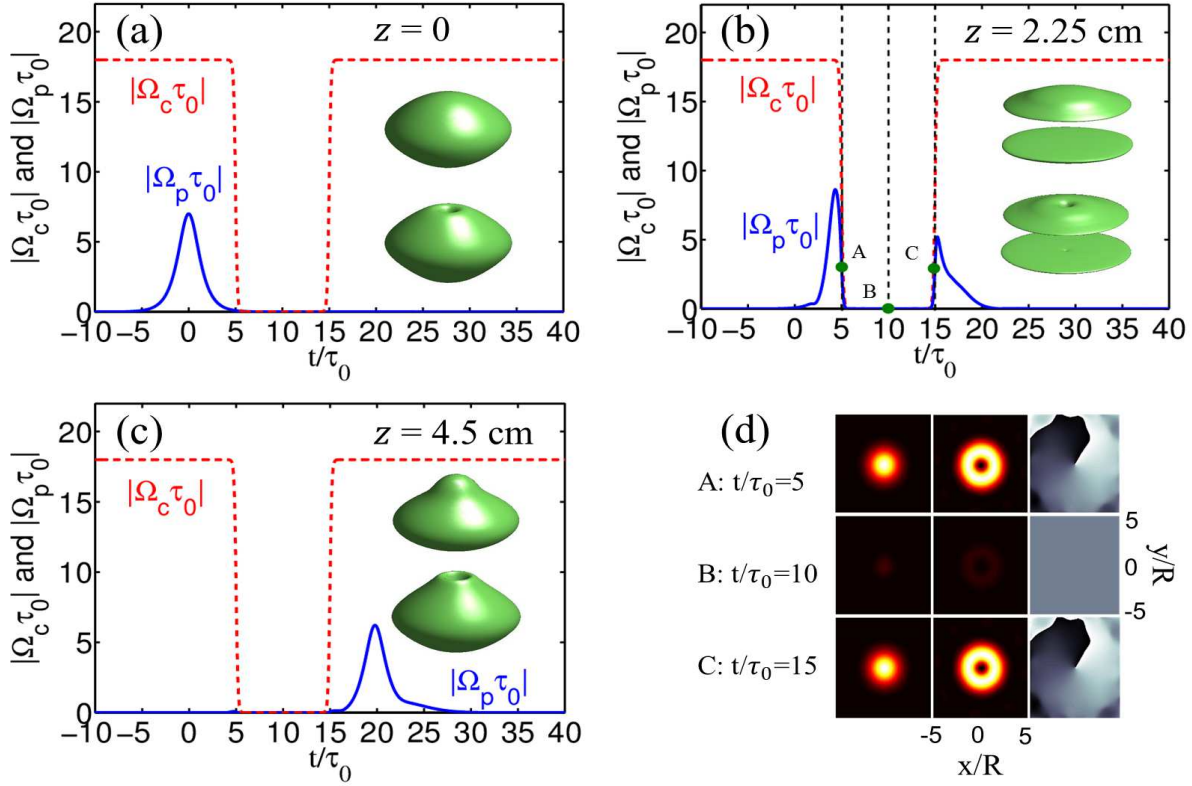}
\caption{\footnotesize {\bf Robust storage and retrieval of (3+1)D light bullet and vortex.} (a), (b), (c) Evolutions of $|\Omega_p\tau_0|$ as function of $t$ at the position $z=0$, $z=2.25$ cm, and $z=4.5$ cm, respectively. Insets are isosurface plots ($|\Omega_p\tau_0|=0.5$) of the light bullet and vortex.
(d) Light-intensity distribution of the light bullet (the first column) and the vortex (the second column) in $x$-$y$ plane at $t/\tau_0=$ 5.0, 10.0, and 15.0, respectively. The third column shows the phase distribution of the vortex. The points $A, B, C$ in (b) correspond to the times $t/\tau_0=5.0, 10.0, 15.0$  in (d). }\label{fig:5}
\end{figure}

In order to illustrate more clearly the evolution process of the storage and retrieval of the stable (3+1)D light bullet and vortex (i.e., the case (ii) described above), in Fig.~\ref{fig:5}(a), Fig.~\ref{fig:5}(b), and Fig.~\ref{fig:5}(c) we show the numerical result of the evolution of the probe field ($|\Omega_p\tau_0|$) and the control field ($|\Omega_c\tau_0|$) as functions of time at $z=0$, 2.25 cm, and 4.5 cm, respectively. We see that the light bullet and vortex undergo steps of appearance, disappearance, and reappearance. Presented in the first (second) column of Fig.~\ref{fig:5}(d) is the light-intensity distribution in $x$-$y$ plane of the bullet (vortex) for $t/\tau_0=$ 5.0, 10.0, and 15.0, respectively. The third column is the phase distribution of the light vortex. The result shows that the light bullet and vortex can be stored around $t/\tau_0=5.0$ when the control field is switched off, and be retrieved around $t/\tau_0=15.0$ when the control field is switched on again. Interestingly, the phase distribution of the vortex can also be stored and retrieved, which means that the memory of the light vortex can bring more information than that of the light bullet.

%
%

We have also studied the storage and retrieval of vortices for $m=2$. The numerical result shows that these
vortices are unstable during the propagation, and hence a robust storage and retrieval of them are
not available.

\vspace{6mm}
\noindent\textbf{\Large \sf Discussion}

From the results described above, a robust SLR for the (3+1)D weak-light bullets and vortices is possible
by using the cold $\Lambda$-type three-level atomic system. These results can be easily generalized to other types of EIT systems with different (such as ladder-type~\cite{CBH}) level configurations. Furthermore, our theory can also be used to study the (3+1)D LSR with a Raman scheme~\cite{Gor,Nunn}, which has been suggested to obtain a broadband quantum memory of linear light pulses and has been realized recently by experiment by using the atomic ensemble working at room temperature~\cite{Reim,Spar}.

In conclusion, we have proposed an EIT-based new scheme to realize a robust LSR for (3+1)D light pulses in a coherent atomic ensemble. Based on MB equations we have derived a nonlinear equation controlling the evolution of the probe-field envelope. We have shown that it is possible to obtain (3+1)D light bullets and vortices, which have very slow propagating velocity and ultra low generation power. We have further shown that these high-dimensional light pulses can be stabilized by using the balance between dispersion, diffraction, nonlinearity, and by a Stark laser field. We have demonstrated that these high-dimensional light pulses can be stored and retrieved very stably by switching off and on a control field. Our study raise the possibility of guiding a related experiment and have potential applications in the area of light and quantum information processing.

\vspace{5mm}
\noindent\textbf{\Large \sf Methods}\\
{\small
\noindent\textbf{ Maxell-Bloch equations}.
In our semi-classical approach, MB equations are used to describe the motion of light field
and atoms. Explicit expressions of the Bloch equation in the interaction picture are
\begin{subequations}
\begin{eqnarray}
&&i\frac{\partial}{\partial
t}\sigma_{11}-i\Gamma_{13}\sigma_{33}+\Omega_p^{\ast}\sigma_{31}-\Omega_p\sigma_{31}^{\ast}=0,\\
&&i\frac{\partial}{\partial
t}\sigma_{22}-i\Gamma_{23}\sigma_{33}+\Omega_{c}^{\ast}\sigma_{32}-\Omega_{c}\sigma_{32}^{\ast}=0,\\
&&i\frac{\partial}{\partial
t}\sigma_{33}+i(\Gamma_{13}+\Gamma_{23})\sigma_{33}-\Omega_p^{\ast}\sigma_{31}+\Omega_p\sigma_{31}^{\ast}-\Omega_c^{\ast}\sigma_{32}+\Omega_c\sigma_{32}^{\ast}=0,\\
&&\left(i\frac{\partial}{\partial
t}+d_{21}\right)\sigma_{21}-\Omega_p\sigma_{32}^{\ast}+\Omega_c^{\ast}\sigma_{31}=0,\\
&&\left(i\frac{\partial}{\partial
t}+d_{31}\right)\sigma_{31}-\Omega_p(\sigma_{33}-\sigma_{11})+\Omega_c\sigma_{21}=0,\\
&&\left(i\frac{\partial}{\partial
t}+d_{32}\right)\sigma_{32}-\Omega_c(\sigma_{33}-\sigma_{22})+\Omega_p\sigma_{21}^{\ast}=0,
\end{eqnarray}
\end{subequations}
where $d_{jl}=\Delta_{j}^\prime-\Delta_{l}^\prime+i\gamma_{jl}$.
Dephasing rates are defined as
$\gamma_{jl}=(\Gamma_j+\Gamma_l)/2+\gamma_{jl}^{\rm col}$ with
$\Gamma_j=\sum_{E_i<E_j}\Gamma_{ij}$  being the spontaneous
emission rate from the state $|j\rangle$ to all lower energy
states $|i\rangle$ and $\gamma_{jl}^{\rm col}$  being the
dephasing rate reflecting the loss of phase coherence between
$|j\rangle$ and $|l\rangle$.

\vspace{4mm}
\noindent\textbf{Asymptotic expansion}.
Assume $\sigma_{jl}=\sum_{q=0}^{\infty}\epsilon^q\sigma_{jl}^{(q)}$, with $\sigma_{jl}^{(0)}=\delta_{j1}\delta_{l1}$,
$\Omega_{p}=\sum_{q=1}^{\infty}\epsilon^q \Omega_{p}^{(q)}$, and
$E_s=\epsilon E_s^{(1)}$. Thus $d_{jl}=d_{jl}^{(0)}+\epsilon^{2}d_{jl}^{(2)}$, with
$d_{jl}^{(0)}=\Delta_j-\Delta_l+i\gamma_{jl}$ and $d_{jl}^{(2)}=
\frac{\alpha_j-\alpha_l}{2\hbar}|E_s^{(1)}|^2$. Here $\epsilon$ is the dimensionless small parameter characterizing the typical amplitude of the probe pulse. To obtain a divergence-free expansion, all the quantities on the right-hand side of the expansion are considered as functions of the multi-scale variables $x_1=\epsilon x$, $y_1=\epsilon y$, $z_{q}=\epsilon^{q}z$ ($q=0,\,1,\,2$), and $t_{q}=\epsilon^{q}t$ ($q=0,\,1$). Substituting the expansions into Eqs.~(\ref{DME}) and~(\ref{MEa}) and comparing the coefficients of $\epsilon^q$, we obtain a set of linear but inhomogeneous equations which can be solved order by order.

The first order ($q=1$) solution is given by $\Omega_{p}^{(1)}=Fe^{i\theta}$ and
$\sigma_{j1}^{(1)}=\{[\delta_{j3}(\omega+\Delta_2+i\gamma_{21})-\delta_{j2}\Omega_c^\ast]/D\}Fe^{i\theta}$,
where $D=|\Omega_c|^2-(\omega+\Delta_2+i\gamma_{21})(\omega+\Delta_3+i\gamma_{31})$  and
$\theta=K(\omega)z_0-\omega t_0$. The linear dispersion relation reads $K(\omega)=\omega/c+\kappa_{13}(\omega+\Delta_2+i\gamma_{21})/D$. $F$ is a yet to be determined envelope function depending on the slow variables $x_1$, $y_1$, $t_1$, $z_1$, and $z_2$.

At the second order ($q=2$), a solvability condition gives  i[$\partial F/\partial z_{1}+(\partial K/\partial \omega)\partial F/\partial t_{1}]=0$, with $V_g=(\partial K/\partial \omega)^{-1}$.
The approximation solution at this order reads
$\sigma_{21}^{(2)}=a_{21}^{(2)}i\frac{\partial}{\partial t_1}Fe^{i\theta}$, $\sigma_{31}^{(2)}=a_{31}^{(2)}i\frac{\partial}{\partial t_1}Fe^{i\theta}$,
$\sigma_{jj}^{(2)}=a_{jj}^{(2)}|F|^2e^{-2\bar{a} z_2}$ ($j=\,1,\,2\,,3$), and
$\sigma_{32}^{(2)}=a_{32}^{(2)}|F|^2e^{-2\bar{a} z_2}$, where
\begin{subequations}
\begin{eqnarray}
&&a_{11}^{(2)}=\frac{\left[i\Gamma_{23}-2|\Omega_c|^2\left(\frac{1}{d_{32}^{(0)}}-\frac{1}{d_{32}^{(0)\ast}}\right)\right]G-i\Gamma_{13}|\Omega_c|^2\left(\frac{1}{Dd_{32}^{(0)\ast}}-\frac{1}{D^\ast
d_{32}^{(0)}}\right)}{i\Gamma_{13}|\Omega_c|^2\left(\frac{1}{d_{32}^{(0)\ast}}-\frac{1}{d_{32}^{(0)}}\right)},\\
&&a_{22}^{(2)}=\frac{G-i\Gamma_{13}a_{11}^{(2)}}{i\Gamma_{13}},\\
&&a_{21}^{(2)}=-\frac{\Omega_c^\ast(2\omega+d_{21}^{(0)}+d_{31}^{(0)})}{D^2},\\
&&a_{31}^{(2)}=\frac{(\omega+d_{21}^{(0)})^2+|\Omega_c|^2}{D^2},\\
&&a_{32}^{(2)}=\frac{\Omega_c}{d_{32}^{(0)}}\left[\frac{1}{D^\ast}-(a_{11}^{(2)}+2a_{22}^{(2)})\right],
\end{eqnarray}
\end{subequations}
and $\bar{a}=\epsilon^{-2}{\rm Im}[K(\omega)]$, with
$G=(\omega+d_{21}^{(0)\ast})/D^\ast-(\omega+d_{21}^{(0)})/D$.

At the third order ($q=3$), a solvability condition yields the equation (\ref{NLSE2}). The explicit expressions of the self- and cross-phase modulation coefficients $W_{11}$ and $W_{12}$ are given by
\begin{subequations}\label{W}
\begin{eqnarray}
&&W_{11}=\kappa_{13}\frac{\Omega_ca_{32}^{(2)\ast}+(\omega+d_{21}^{(0)})(2a_{11}^{(2)}+a_{22}^{(2)})}{D},\\
&&W_{12}=\kappa_{13}\frac{(\omega+d_{21}^{(0)})^2(\alpha_{3}-\alpha_{1})+|\Omega_c|^2(\alpha_{2}-\alpha_{1})}{2\hbar
D ^2}.
\end{eqnarray}
\end{subequations}
}


\vspace{5mm}
\noindent\textbf{\Large \sf Acknowledgments}\\
\noindent This work was supported by the National Natural Science Foundation of China under Grants No. 11174080, 11105052, and 11204274.

\vspace{5mm}
\noindent\textbf{\Large \sf Author contributions}\\
\noindent Z.C. carried out the analytical and numerical calculations. Z.B., H.-j.L. and C.H. developed primary calculating code and helped the numerical calculation. Z.C. and C.H. wrote the manuscript.  G.H. conceived the idea, conducted the calculation and revised the manuscript.

\vspace{5mm}
\noindent\textbf{\Large \sf Additional information}\\
Competing financial interests: The authors declare no competing financial interests.

\end{document}